\documentclass[12pt,fleqn]{article}             
\newcounter{saveeqn}
\usepackage{epsfig}

\usepackage{amsfonts}          
\usepackage{amssymb}           
\usepackage{amsmath}           

 
\def\references{\section*{References}
    \bgroup\parindent=1cm\parskip=1mm\small
    \def\refpar{\par\hangindent=1em\hangafter=1}\reference\ \vskip-10mm}
\def\endreferences{\refpar\egroup}

\def\reference{\relax\refpar \ \hskip-1.1cm} 

\renewcommand{\vec}[1]{\mbox{\boldmath $ #1$}}

\newcommand{\captionfonts}{\small \sl}
\makeatletter  
\long\def\@makecaption#1#2{%
  \vskip\abovecaptionskip
  \sbox\@tempboxa{{{\bf #1:} \captionfonts #2}}%
  \ifdim \wd\@tempboxa >\hsize
    { {\bf#1:} \captionfonts#2\par}
  \else
    \hbox to\hsize{\hfil\box\@tempboxa\hfil}%
  \fi
  \vskip\belowcaptionskip}
\makeatother   

\newcommand{\kv}{{\bf k}} 
 
\newcommand{\uv}{{\bf u}}

\title{\bf Convection in Rotating Spherical Fluid Shells}
\author{\sf F. H. Busse\footnote{e-mail: {\tt Busse@uni-bayreuth.de}}
  \  and R. Simitev}
\date{\small Institute of Physics, University of Bayreuth, D-95440
  Bayreuth, Germany}

\begin{document}
\maketitle

"Mathematical Aspects of Natural Dynamos", E. Dormy, A.M. Soward
(eds.), Grenoble Sciences and CRC Press, Boca Raton, pp. 119-198,
ISBN-13:978-1-58488-954-0, 2007.

\section{Introduction}

Convection driven by thermal buoyancy in rotating spherical bodies of
fluid has long been  recognized as a fundamental process in the
understanding of the properties of planets and stars. Since these
objects are rotating in general and since their evolution is
associated with the transport of heat from their interiors convection
influenced by the Coriolis force does indeed play a dominant role in
the dynamics of their fluid parts. In the case of the Earth it is the
generation  of the geomagnetic field by motions in the molten outer
iron core which has stimulated much interest in the subject of
convection in rotating spheres. But the zones and belts seen on
Jupiter  are a just as interesting phenomenon driven by convection in
the deep atmosphere of the planet. Similarly, the differential
rotation of the sun and its magnetic cycle are intimately connected
with the solar convection zone encompassing the outer 29 percent of
the Sun in terms of its radial extent.

Theoretical studies of convection in rotating fluid spheres started
about 50 years ago. The attention was restricted to the linear problem
of the onset of convection and for simplicity axisymmetric motions
were assumed. An account of these early efforts can be found in
Chandrasekhar's famous treatise (1961) and in the papers by Bisshopp
and Niiler (1965) and by Roberts (1965). A little later it became
evident that the preferred forms of convection in the interesting
limit of rapid rotation are not axisymmetric, but highly
non-axisymmetric (Roberts, 1968). In this later paper, however, the
incorrect assumption was made that the preferred mode of convection
exhibits a $z$-component  of the velocity field parallel to the axis
that is symmetric with respect to the equatorial plane. The correct
mode for the onset of convection was found by Busse (1970a) who
approached the problem on the basis of the rotating cylindrical
annulus model. This model takes advantage   of the approximate validity of the 
Proudman-Taylor theorem and the analysis can thus be reduced from three to two
spatial dimensions. We shall devote section 2 to a description of this
model since it offers the simplest access to the spherical problem.The
basic equations for the spherical problem are introduced in section 3
and the onset of columnar convection in spherical shells is discussed
in section 4. In section 5 the onset of inertial mode convection is
described which prevails at very low Prandtl numbers. In section 6 the
properties of finite amplitude convection are outlined for moderately
low Prandtl numbers $Pr$, while convection at higher values of $Pr$ is
considered in section 7. In section 8 equatorially attached convection
is considered which evolves from inertial mode convection. The problem
of penetrative convection is addressed in section 9 and in section 10
some aspects of convection in the presence of thermal as well as
chemical  buoyancy are discussed. The chapter concludes with some
remarks on applications among which the dynamo action of convection is
of special importance.


\section{ Convection in the Rotating Cylindrical Annulus}

\begin{figure}[t]
\begin{center}
\epsfig{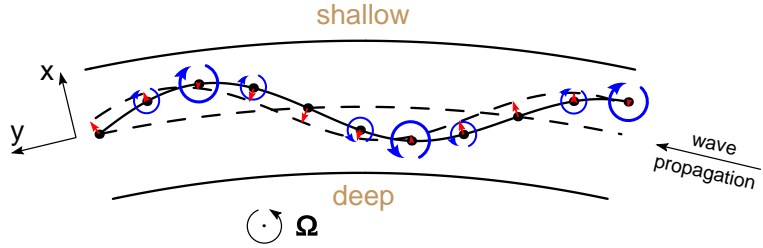}
\end{center}
\caption{The mechanism of propagation of a Rossby wave visualized in the
equatorial plane of the rotating annulus: Fluid columns originally resting 
at the mid-surface acquire anti-cyclonic vorticity relative to the rotating
system when they are displaced outwards towards the shallow region.
Cyclonic vorticity is acquired by the columns displaced inwards.
The action of the columnar motion on the neighboring fluid columns is such
that an initial sinusoidal displacement propagates in the prograde direction.}
\label{f2}
\end{figure}

Convection in the fluid filled gap between two rigidly rotating
coaxial cylinders is receiving increasing attention because it shares
many linear and nonlinear dynamical properties with convection in
rotating spherical fluid shells, at least as far as columnar
convection is concerned. 
\begin{figure}[t]
\begin{center}
\hspace*{-1cm}\epsfig{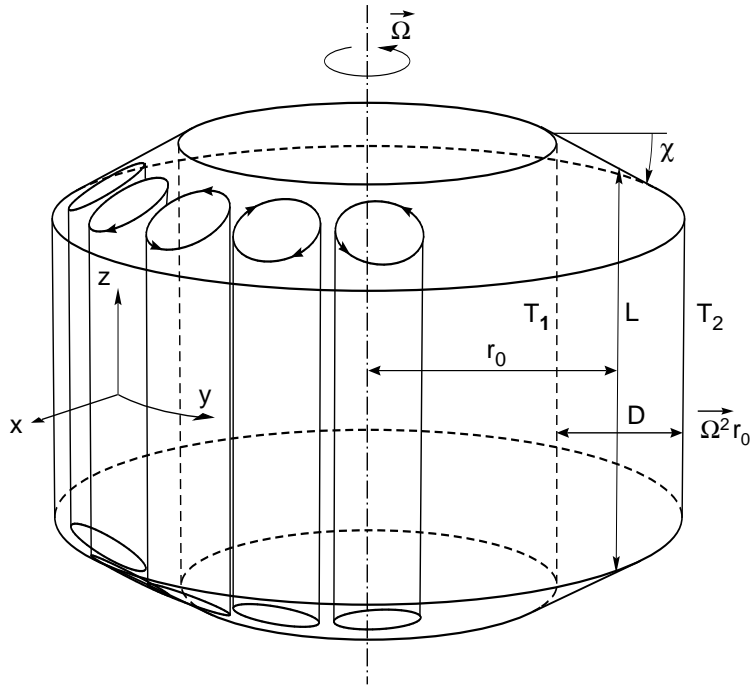}
\end{center}
\caption{Sketch of the geometrical configuration of the rotating
  cylindrical annulus.}
\label{f1}
\end{figure}
From the point of view of planetary
applications it seems natural to have gravity pointing inward and to
keep the inner cylinder at the higher temperature $T_2$. But since the
centrifugal force is used as a source of buoyancy in laboratory
experiments with the higher temperature at the outer cylinder we shall
use this latter configuration as shown in Fig.~\ref{f1}. Since only
the product of effective gravity and applied temperature gradient is
physically relevant the two cases are equivalent. In experimental
realizations of the system  ordinary gravity plays a minimal role when
a vertical axis of rotation is used and when the rate of rotation is
sufficiently high such that the centrifugal force exceeds gravity by
at least a factor of two or three.\ An important ingredient of the
geometrical configuration shown  in Fig.~\ref{f1}  are the conical
boundaries at top and bottom which cause a variation in height with
distance from the axis of rotation. Without this variation in height
steady two-dimensional convection rolls aligned with the axis will be
realized since they obey the Proudman-Taylor condition. The Coriolis
force is entirely balanced by the pressure  gradient in this case and
Rayleigh number for onset of convection in the small gap limit is
given by the Rayleigh-B\'enard value for a non-rotating layer. 
Thin Ekman layers at the no-slip top and bottom boundaries exert only a minor
influence on the dynamics of convection if the height $L$ of the annulus 
is sufficiently large in comparison to the gap size.
As soon as the height changes in the radial direction any flow involving
a radial velocity component can no longer satisfy the geostrophic balance. 
Instead a weak time dependence is required and the flow assumes the
character of Rossby waves.These waves are well known in the
meteorological context where the variation of the vertical component
of rotation with latitude has the same effect as  variation of height
in the annulus of Fig.~\ref{f1}. The dynamics of Rossby waves can be
visualized most readily if the action of the vorticity acquired by the
fluid columns displaced radially from the middle of the gap   is
considered. As indicated in Fig.~\ref{f2} columns shifted inward
acquire cyclonic vorticity  because they are stretched owing to the
increasing height. The opposite sign of vorticity is exhibited by
columns moving outward. Since  their moments of inertia are increased
they must rotate anti-cyclonically  relative to the rotating system in
order to conserve angular momentum. The action of the acquired motion
of sinusoidally displaced columns on their neighbors results in the
propagation of a wave as shown in Fig.~\ref{f2}. The phase velocity
is in the prograde (retrograde) direction  hen the height decreases
(increases) with distance from the axis.

In the case of convection the dynamics is modified by the presence of
thermal buoyancy and the phase velocity is less than that of Rossby
waves except in the limit of vanishing Prandtl number. The analysis of
the thermal Rossby waves, as the propagating convection waves are
called, is quite simple if the inclination of the cones with respect
to the equatorial plane is introduced as a small perturbation. Using
the gap width $D$ as length scale and $D^2 / \nu$ as time scale where
$\nu$ is the kinematic viscosity of the fluid we may assume the
velocity field in the form 
\begin{equation}
\uv = \nabla \psi (x,y,t) \times \kv + \ldots
\end{equation}
where $\kv$ is the unit vector in the $z$-direction parallel to the axis
of rotation and where the small gap approximation with $x$ as radial coordinate
and $y$ as azimuthal coordinate has been assumed. Only the geostrophic
component of $\uv$ has been denoted explicitly in expression (1).
Deviations from geostrophy are induced by the condition of vanishing
normal velocity at the conical boundaries,
\begin{equation}
u_x \eta_0 \pm u_z = 0 \quad \mbox{ at } \quad z = \pm \frac{L}{2D}
\end{equation}
where the tangent $\eta_0$ of the angle $\chi$ between the cones and
the equatorial plane has been introduced as small parameter.
By taking the $z$-component of the curl of the equation of motion and
averaging it over the height of the annulus we can incorporate the
boundary condition (2) into an equation for $- \Delta_2 \psi$ which is
the $z$-component of vorticity, 
\begin{equation}
\left( \frac{\partial}{\partial t} + \frac{\partial}{\partial y} \psi
\frac{\partial}{\partial x} - \frac{\partial}{\partial x} \psi 
\frac{\partial}{\partial y} \right) \Delta_2 \psi - \Delta_2^2 \psi - 
\eta \frac{\partial}{\partial y} \psi = - \frac{\partial}{\partial y} \Theta
\end{equation}
where $\Delta_2$ is the two-dimensional Laplacian, $\Delta_2 = \partial^2 /
\partial x^2 + \partial^2 / \partial y^2$. Equation (3) must be
considered together with the heat equation for the deviation $\Theta$
from the static temperature distribution of pure conduction,
\begin{equation}
Pr \left( \frac{\partial}{\partial t} + \frac{\partial}{\partial y} \psi 
\frac{\partial}{\partial x} - \frac{\partial}{\partial x} \psi 
\frac{\partial}{\partial y} \right) \Theta + \frac{\partial}{\partial y}
\psi = \Delta_2 \Theta .
\end{equation}
where $\Theta$ is measured in multiples of $(T_2 - T_1) Pr/Ra$. $T_1$
and $T_2$ are the temperatures prescribed at the inner and outer
cylindrical boundaries, respectively. The dimensionless Coriolis
parameter $\eta$, the Rayleigh number $Ra$ and  the Prandtl number
$Pr$ are defined by 
\begin{displaymath}
\eta = \frac{4 \Omega \eta_0 D^3}{\nu L} , \quad Ra = \frac{\gamma (T_2 - T_1)
\Omega ^2 r_0 D^3}{\nu \kappa} , \quad Pr = \frac{\nu}{\kappa} .
\end{displaymath}
where $\Omega$ denotes the angular velocity of rotation, $\gamma$ is the
coefficient of thermal expansion, $\kappa$ is the thermal diffusivity
and $r_0$ is the mean radius of the annulus. Assuming stress-free
boundaries at the cylindrical walls,  
\begin{equation}
\psi = \frac{\partial^2}{\partial x^2} \psi = \Theta = 0 \quad \mbox{ at }
\quad x = \pm \frac{1}{2},
\end{equation}
we obtain a completely specified mathematical formulation of the
problem of centrifugally driven convection in the cylindrical
annulus. 

The onset of convection is described by the linearized version of
Eqs.~(3) which can be solved by
\begin{align}
&\psi = A \sin n \pi ( x + \frac{1}{2} ) \exp \{ i \alpha y + i \omega
t ), \nonumber \\
&\Theta = \frac{- i \alpha \psi}{\alpha^2 + ( n \pi)^2 + i \omega}, 
\end{align}
with the following relationships for $\omega$ and $Ra$
\begin{align}
&\omega = \frac{- \eta \alpha}{(1+Pr)(n^2\pi^2 + \alpha^2)} , \nonumber\\
&Ra = (n^2\pi^2 + \alpha^2 )^3 \alpha^{-2} + \left( \frac{\eta Pr}{1+Pr} 
\right)^2 (n^2\pi^2 + \alpha^2)^{-1} .
\end{align}
As expected the dependence of Rayleigh number on the wavenumber in the
case of Rayleigh-B\'enard convection in a non-rotating layer is
recovered in the  limit $\eta = 0$.
The mode corresponding to $n=1$ is preferred in this case, of
course. This property continues to hold for finite $\eta$. But as the
limit $\eta \rightarrow \infty$ is approached, the values of $Ra$ and
$\omega$ do not depend on $n$ in first approximation as can be seen  
from the following expressions for the critical values in the limit of
large $\eta$
\begin{align}
&\alpha_c = \eta_P^{\frac{1}{3}} ( 1 - \frac{7}{12} \pi^2 \eta_P^{-\frac{2}{3}}
+ \ldots ) \nonumber \\
&Ra_c = \eta_P^{\frac{4}{3}} ( 3 + \pi^2 \eta_P^{-\frac{2}{3}}
+ \ldots ) \nonumber \\
&\omega_c = - \sqrt{2} Pr^{-1} \eta_P^{\frac{2}{3}} 
(1- \frac{5}{12} \pi^2 \eta_P^{-\frac{2}{3}} + \ldots )
\end{align}
where the definition $\eta_P = \eta Pr  \sqrt{1/2} (1 + Pr)^{-1}$ has
been used. Expressions (8) have been derived for $n=1$, but they hold
for arbitrary $n$ when $\pi^2$ is replaced by $(n \pi)^2$ The weak
dependence on the radial coordinate of the problem has two important
consequences: 
\begin{itemize}
\item[(i)]
The onset of convection is rather insensitive to the cylindrical
boundaries. The analysis can thus be applied to the case of a sphere
where these boundaries are missing.
\item[(ii)]
Modes of different radial dependence correspond to the same critical
parameters asymptotically. Secondary bifurcations right above
threshold become possible through couplings of these modes.
\end{itemize}
The latter possibility is indeed realized in the form of the mean flow
instability. A transition to a solution of the form 
\begin{equation}
\hspace*{-1.5cm}\psi = A \sin ( \alpha y - \omega t ) \sin \pi ( x + \frac{1}{2} ) + B
\sin ( \alpha y - \omega t + \varphi ) \sin 2 \pi ( x + \frac{1}{2} )
\end{equation}
occurs as the Rayleigh number is increased beyond the critical value
unless the Prandtl number is rather small (Or and Busse, 1987). A
characteristic property of solution (9) is the strong mean zonal shear
which it generates through its Reynolds stress, $\overline{u_x u_y}
\propto AB \sin \pi  (x + \frac{1}{2} ) \sin \varphi $, where the bar
indicates the average over the $y$-coordinate. Both signs of the shear
are equally possible since the sign of $B$ is arbitrary.
\begin{figure}[t]
\begin{center}
\epsfig{file=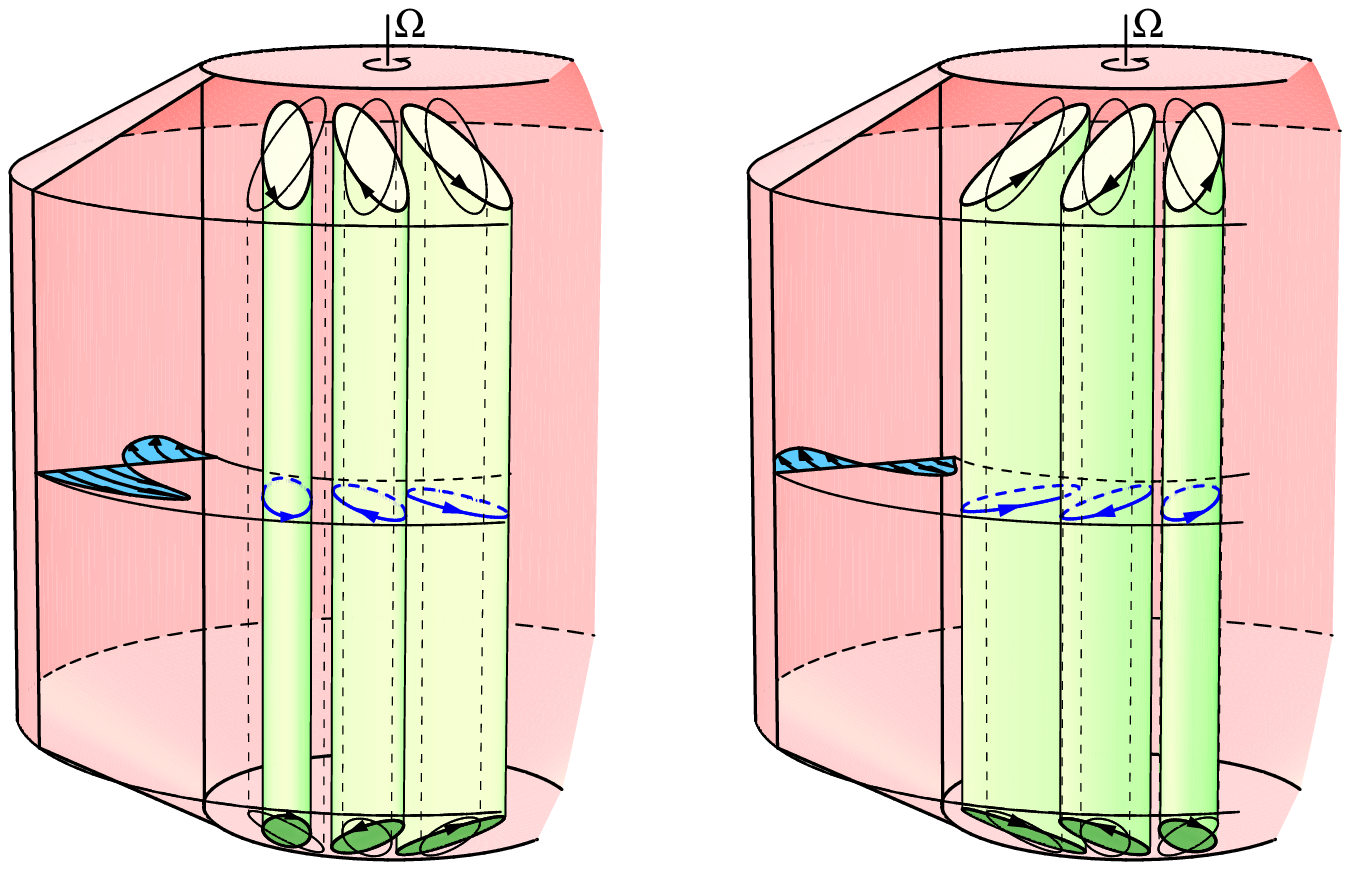,width=10.cm}
\end{center}
\caption{The mean flow instability leading to either outward (right
  figure) or inward (left figure) transport of prograde angular
  momentum.} 
\label{f3}
\begin{center}
\epsfig{file=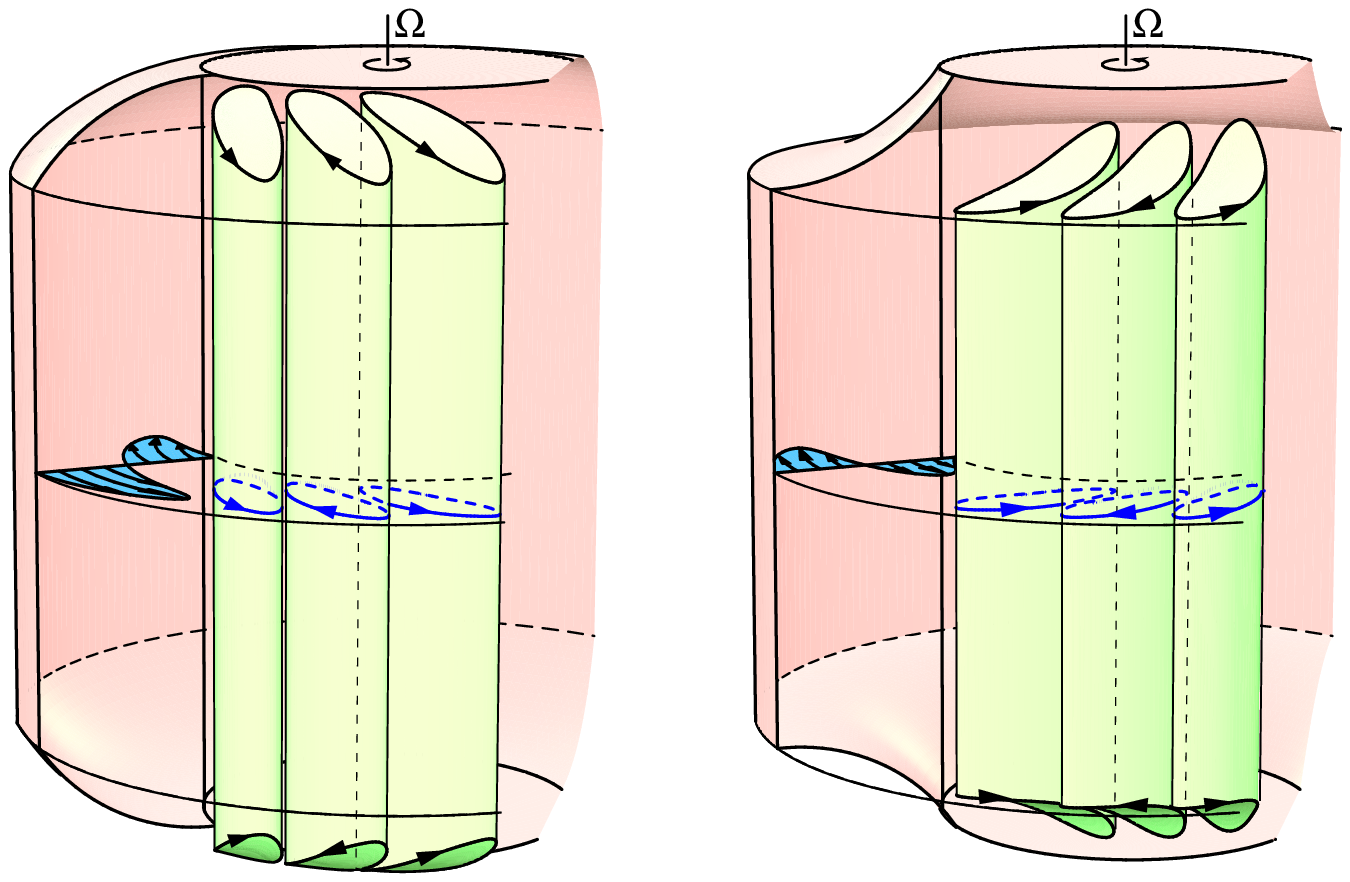,width=10.cm}
\end{center}
\caption{Influence of curved conical boundaries.  In the convex case
  (right figure) the columns tend to spiral outward in the  prograde
  direction. They thus create a different rotation with higher angular
  velocity on the outside than on the inside. The opposite situation
  is found in the concave case (left figure). }
\label{f4}
\end{figure}
The mean flow instability corresponds to a tilt of the convection
columns as indicated in Fig.\ \ref{f3}. When the columns are slightly
tilted in the prograde sense towards the outside prograde momentum is
carried outward and retrograde momentum is transported inwards leading
to a differential rotation in which the outer fluid rotates faster
than the inner one. An equilibrium is reached through viscous stresses
which tend to oppose the differential rotation. The reverse situation
occurs when the columns are tilted the other way as  shown in the
right plot of Fig.\ \ref{f3}. The instability occurs because the
differential rotation tends to increase the initial tilt and a
feedback process is thus initiated. The mean flow instability of
convection rolls is also possible in a  non-rotating
Rayleigh-B\'enard layer. But there it is usually preceded by
three-dimensional instabilities.

There is another way in which a differential rotation in the annulus can 
be generated. When curved cones instead of straight cones are used as indicated in 
Fig.~\ref{f4} solutions of the form (5) with separating $x$- and
$y$-dependences are no longer possible. The term $\eta \partial_y
\psi$ in Eq. (3a) must now be replaced  by $\eta ( 1 + \epsilon x)
\partial_y \psi$ where positive $\epsilon$ refers to the convex case
of Fig.~\ref{f4}  while a negative $\epsilon$ corresponds to the
concave cones of the right plot. The columns are tilted  because the
thermal Rossby wave has the tendency to propagate faster on the
outside than on the inside when $\epsilon$ is positive and vice versa.
A differential rotation prograde on the outside and retrograde on
the inside must thus be expected for $\epsilon > 0$ as shown in 
the left  plot of  Fig.~\ref{f4} while the opposite results is
obtained for $\epsilon < 0$. The experiment of Busse and Hood (1982)
has demonstrated this effect.

There are numerous other interesting features of convection in the
cylindrical  annulus such as vacillations and relaxation oscillations
which appear at higher Rayleigh numbers and which can be related to
analogous phenomena of convection in rotating spheres. We refer to the
papers by Or and Busse (1987), Schnaubelt and Busse (1992),  Brummell
and Hart (1993) for details. The influence of non-axisymmetric
modulations of the boundaries can also easily be investigated in the
annulus model as shown by  Bell and Soward (1996), Herrmann and Busse
(1997) and Westerburg and Busse (2003). Besides the narrow gap limit
the finite gap case of the rotating cylindrical annulus system is also
of interest. The discrete manifold of realizable wavenumbers gives
rise to resonances and changes in the character   of secondary
instabilities. We refer to the recent papers by Pino {\it et al.}
(2000, 2001) and Chen and Zhang (2002).  


\section{Mathematical Formulation of the Problem of Convection in
  Rotating Spherical Shells} 

The presence of the centrifugal force in rotating planets and stars
usually causes some deviations from spherical symmetry. The surfaces
of constant potential become spheroidal and a basic state of vanishing
motion relative to a rotating frame of reference may not exist since
surfaces of constant density do not coincide with surfaces of constant
potential in general. As a result of this baroclinicity a differential
rotation must be expected (see, for example, Busse, 1982). But these
effects are usually much smaller than those introduced by convection
and it is thus a good approximation to neglect the effects of the
centrifugal force and to assume that there exists a basic static
solution with spherically symmetric distributions of gravity and
temperature. 
 
For the theoretical description of thermal convection in rotating
spheres usually the case of a gravity varying linearly with radius,
${\bf g} = - g_0 d {\bf r}$ is assumed where ${\bf r}$ is the position
vector with respect to  the center of the sphere which is made
dimensionless with the thickness $d$ of the shell. We assume that a
static state exists with the temperature distribution $T_S = T_0 -
\beta  d^2 r^2 /2 + \Delta T \eta r^{-1} (1-\eta)^{-2}$ where $\eta$
denotes the ratio of inner to outer radius of the shell and $\beta$ is
proportional to a uniform density of heat sources in the sphere. In
addition an applied temperature difference is admitted such that  
$\Delta T$ is the temperature difference between the boundaries in the
special case $\beta =0$. Of course, in geophysical and astrophysical
applications only the super-adiabatic part of the temperature field
must be identified with the temperature distribution given above.

In addition to  $d$, the time $d^2 / \nu$ and the temperature $\nu^2 /
g_0 \gamma d^4$  are used as scales for the dimensionless description
of the problem  where $\nu$ denotes the kinematic viscosity of the
fluid and $\kappa$ its thermal diffusivity. We use the Boussinesq
approximation in that we assume the density $\rho$ to be constant
except in the gravity term where its temperature dependence given by
$\gamma \equiv ( d \varrho/dT)/\varrho =$ const. is taken into
account.  The basic equations of motion and the heat equation for the
deviation $\Theta$ from the static temperature distribution are thus
given by 
\begin{subequations}
\begin{align}
\label{1a}
&\hspace*{-1cm}\partial_t \vec{u} + \vec u \cdot \nabla \vec u + \tau \vec k \times
\vec u = - \nabla \pi +\Theta \vec r + \nabla^2 \vec u \\
\label{1b}
&\hspace*{-1cm}\nabla \cdot \vec u = 0 \\
\label{1c}
&\hspace*{-1cm} Pr (\partial_t \Theta + \vec u \cdot \nabla \Theta) = (Ra_i+Ra_e\eta
r^{-3}(1-\eta)^{-2}) \vec r \cdot \vec u + \nabla^2 \Theta 
\end{align}
\end{subequations}
where the Rayleigh numbers $Ra_i$ and $Ra_e$, the Coriolis number
$\tau$ and  the Prandtl number $Pr$ are defined by  
\begin{displaymath}
\label{1d}
Ra_i = \frac{\gamma g_0 \beta d^6}{\nu \kappa} ,  \enspace 
Ra_e =\frac{\gamma g_0 \Delta T d^4}{\nu \kappa} , \enspace \tau =
\frac{2 \Omega d^2}{\nu} , \enspace 
Pr = \frac{\nu}{\kappa}. 
\end{displaymath}
 Since
the velocity field $\vec u$ is  solenoidal the general representation
in terms of poloidal and toroidal components can be used, 
\begin{displaymath}
\label{1e}
\vec u = \nabla \times ( \nabla v \times \vec r) + \nabla w \times
\vec r .
\end{displaymath}
By multiplying the (curl)$^2$ and the curl of equation (\ref{1a}) by
$\vec r$ we obtain two equations for $v$ and $w$,  
\begin{subequations}
\begin{align}
\label{2a}
&\hspace*{-1.8cm}[( \nabla^2 - \partial_t) L_2 + \tau \partial_{\phi} ] \nabla^2 v +
\tau Q w - L_2 \Theta  = - \vec r \cdot \nabla \times [ \nabla \times
( \vec u \cdot \nabla \vec u )] \\
\label{2b}
&\hspace*{-1.8cm}[( \nabla^2 - \partial_t) L_2 + \tau \partial_{\phi} ] w - \tau Qv
= \vec r \cdot \nabla \times ( \vec u \cdot \nabla \vec u), 
\end{align}
\end{subequations}
where $\partial_t$ and $\partial_{\phi}$ denote the partial
derivatives with respect to time $t$ and with respect to the  angle
$\phi$ of a spherical system of coordinates $r, \theta, \phi$
and where the operators $L_2$ and $Q$ are defined by  
\begin{subequations}
\begin{align}
\label{3a}
&L_2 \equiv - r^2 \nabla^2 + \partial_r ( r^2 \partial_r)\\
\label{3b}
&Q \equiv r \cos \theta \nabla^2 - (L_2 + r \partial_r ) ( \cos \theta
\partial_r - r^{-1} \sin \theta \partial_{\theta}).  
\end{align}
\end{subequations}
Stress-free boundaries with fixed temperatures are most often assumed,
\begin{gather}
\label{4}
v = \partial^2_{rr}v = \partial_r (w/r) = \Theta = 0 \\
\mbox{at } \enspace r=r_i \equiv \eta / (1- \eta) \enspace \mbox{ and}
\enspace r=r_o = (1-\eta)^{-1}. \nonumber
\end{gather}
The numerical integration of equations (2) together with boundary
conditions (4) in the general nonlinear case proceeds with the
pseudo-spectral  method as described by Tilgner and Busse (1997)
which is based on an expansion of all dependent variables in
spherical harmonics for the $\theta , \phi$-dependences, i.e. 
\begin{equation}
v = \sum \limits_{l,m} V_l^m (r,t) P_l^m ( \cos \theta ) \exp \{ im \phi \}
\end{equation}
and analogous expressions for the other variables, $w$ and $\Theta$. 
$P_l^m$ denotes the associated Legendre functions.
For the $r$-dependence expansions in Chebychev polynomials are used. 
For further details see also Busse {\it et al.} (1998).
For the computations to be reported in sections 5 and 6 a minimum of
33 collocation points in the radial direction and spherical harmonics
up to the order 64 have been used. But in many cases the resolution
was increased to 49 collocation points and spherical harmonics up to
the order 96 or 128. 


\section{The Onset of Convection in Rotating Spherical Shells}

\begin{figure}[t]
\begin{center}
\epsfig{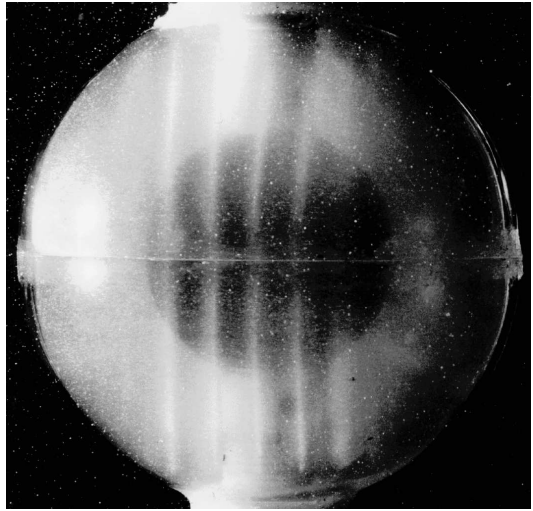}
\end{center}
\caption{Centrifugally driven convection in a rotating fluid between
  an inner cooled and an outer heated spherical boundary. The motions
  are visualized through nearly neutrally buoyant thin platelets which
  align themselves with the shear. The photograph shows the onset of
  thermal Rossby waves in the form of convection columns in a thick
  fluid shell.}  
\label{f5}
\end{figure}
The main difficulty in analyzing convection in rotating spherical
shells arises from the fact that the role of the Coriolis force varies
with  the angle between gravity and the vector ${\bf \Omega}$ of angular
velocity. The geometrical configuration of the  polar regions of the
shell thus resembles that of a Rayleigh-B\'enard layer rotating about
a vertical axis while in the equatorial region the model of the
rotating cylindrical annulus can be applied. Only at low rotation
rates does convection set in in a global fashion and an axisymmetric
mode can become preferred in this case (Geiger and Busse, 1981). At
higher rotation rates the onset of convection does indeed occur in the
form of the columnar modes as predicted on the basis of the annulus
model of section 2. A visualisation  of the convection motion in the
form of thermal Rossby waves  traveling in the prograde direction is
shown in Fig.~\ref{f5}. 

\begin{figure}[t]
\begin{center}
\hspace*{0cm}\epsfig{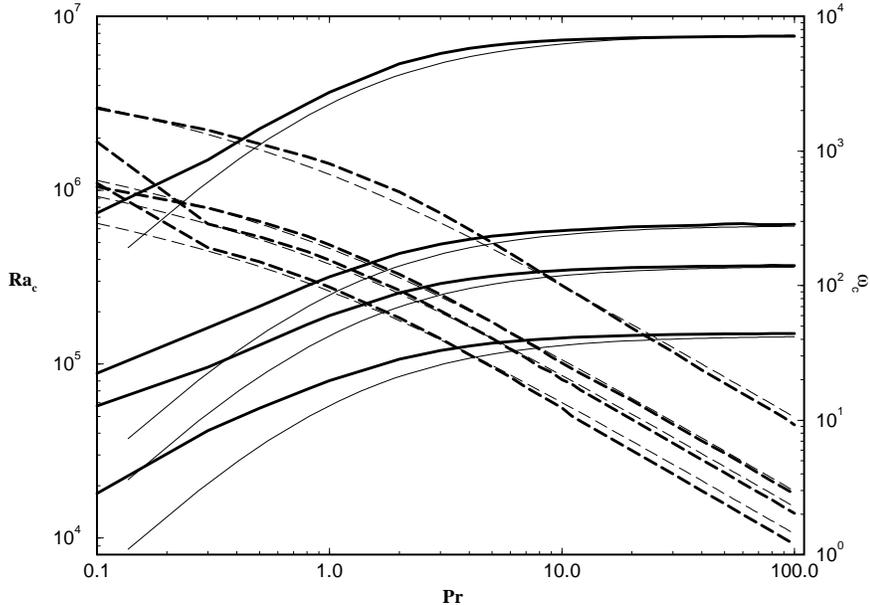}
\end{center}
\caption{Critical Rayleigh number $Ra_{ic}$ (thick lines) and
  frequency $\omega_c$ (right ordinate, thick dashed lines) as a
  function of the Prandtl number $Pr$ in the case $\eta = 0.4$ for the
  Coriolis numbers $\tau = 5 \cdot 10^3$, $10^4$, $1.5 \cdot 10^4$
  and $10^5$. The thick solid line corresponds to expression (15a) and
  (15c).}  
\label{f6}
\end{figure}

A rough idea of the dependence of the critical Rayleigh number $Ra_{ic}$
for the onset of convection on the parameters of the problem in the
case $Ra_e=0$ can be gained from the applications of expressions (8)
of the annulus model, 
\begin{subequations}
\begin{align}
&Ra_{ic} = 3 \left( \frac{Pr \tau }{1+Pr} \right)^{\frac{4}{3}} ( \tan
\theta_m)^{\frac{8}{3}} r_m^{-\frac{1}{3}} 2^{-\frac{2}{3}}   \\
&m_c = \left( \frac{Pr \tau}{1+Pr} \right)^{\frac{1}{3}} ( r_m \tan
\theta_m )^{\frac{2}{3}} 2^{-\frac{1}{6}}  \\ 
&\omega_c = \left( \frac{\tau^2}{(1+Pr)^2Pr} \right)^{\frac{1}{3}}
2^{-\frac{5}{6}}  (\tan^2 \theta_m / r_m )^{\frac{2}{3}},
\end{align}
\end{subequations}
where $r_m$ refers to the mean radius of the fluid shell, $r_m = (r_i
+ r_o)/2$, and $\theta_m$ to the corresponding colatitude, $\theta_m
=$ arcsin $(r_m(1-\eta))$. The azimuthal wavenumber of the preferred
mode is denoted by $m_c$ and the corresponding angular velocity of the
drift of the convection columns in the prograde direction is given by
$\omega_c / m_c$. In figure \ref{f6} the expressions (15a,c) are
compared with accurate numerical values in the case $Ra_e=0$ which
indicate that the general  trend is well represented by expressions
(15a,c). The same property holds for $m_c$. In the case $Ra_i=0$ the
agreement with expressions (15a,c) is not quite as good since the
onset of convection is more concentrated towards the tangent cylinder
touching the inner boundary at its equator because of the higher
temperature gradient in that region. 
Since we shall continue to restrict the attention to the case
$Ra_e=0$, unless indicated otherwise, we shall drop the subscript $i$
of $Ra_i$. 

\begin{figure}[t]
\begin{center}
\hspace*{0cm}\epsfig{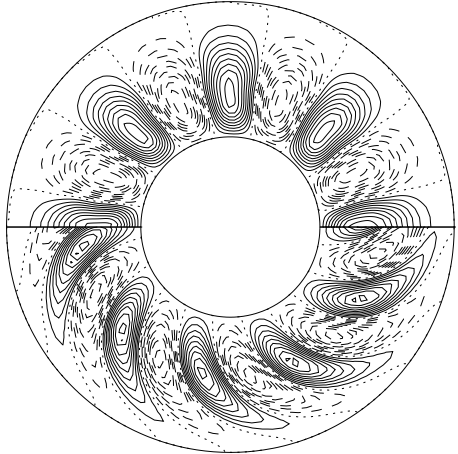}
\end{center}
\caption{Equatorial streamlines, $r \partial v / \partial \varphi =$
  const., of  convection columns in the cases $Pr=10^5$, $\tau=10^5$
  and $Ra=3 \cdot 10^6$  (upper half) and $Pr=1$, $\tau=10^4$ and
  $Ra=2\cdot10^5$  (lower half).} 
\label{f7}
\end{figure}

For the rigorous analysis of the onset problem in the limit of rapid
rotation the dependence of the solution on the distance $s$ from the
axis must be considered which has been neglected in the application of
the annulus model. It turns out that a finite difference exists
between the results of the local analysis and the exact global
analysis as was pointed out already by Soward (1977). Using a WKBJ
approach with the double turning point method of Soward and Jones
(1983), Yano (1992) has analyzed the asymptotic problem in a refined
version of the rotating  cylindrical annulus model. He assumes a
finite gap and the same dependence on $s$ of the small inclination of
the convex conical end surfaces as in the case of the sphere. The
component of gravity parallel to the axis of rotation is still
neglected. These assumptions have been dropped by Jones et al. (2000),
who attacked the full spherical problem. They find that their results
agree surprisingly well with those of  Yano (1992).
The analytical findings have been confirmed through numerous numerical 
studies(Zhang and Busse, 1987; Zhang, 1991, 1992a; Ardes {\it et al.} 1997;
Sun {\it et al.}, 1993) of which we show Fig.~\ref{f7} as an example.This figure
emphasizes the difference between the strongly  spiralling nature of
the convection columns at Prandtl numbers of the order unity or less
and the more radially oriented columns at higher Prandtl numbers. This
property is a result of the strong  decrease of the frequency $\omega$
with increasing $Pr$. 

The interest in convection in rotating spherical shells has motivated
a  number of laboratory investigations of the problem. The experiment
of Hart et al. (1986) in which a spherically symmetric  electric field
acting on a dielectric insulating liquid  is used to simulate gravity was
carried out in space in order to avoid the interference from
laboratory  gravity. Less sophisticated experiments (Busse and
Carrigan, 1976; Carrigan and Busse, 1983; Cardin and Olson, 1994;
Cordero and Busse, 1992; Sumita and Olson, 2000) have used the
centrifugal force  with a cooled inner and a heated outer sphere to
simulate the onset as well as the finite amplitude properties of convection.
The main handicap of these experiments is the zonal flow generated as a 
thermal wind and the associated meridional circulation in the basic
axisymmetric state (Cordero and Busse, 1992). But this handicap can be
minimized through the use of high rotation rates and the observations
correspond quite well to the theoretical expectations as  shown by the
example of Fig.~\ref{f5}. 


\section{Onset of  Inertial Convection at Low Prandtl Numbers}

Onset of instability in the form of inertial convection is well known
from the case of a plane horizontal layer heated from below and
rotating about a vertical axis. As has been discussed by Chandrasekhar
(1961) convection in the form of modified inertial waves represents
the preferred mode at the onset of instability for Prandtl numbers of
the order $0.6$ or less for sufficiently high values of the Coriolis
number $\tau$. A similar situation is found for convection in rotating
spherical shells where Zhang and Busse (1987) identified equatorially
attached modes of convection as modified inertial waves. This
connection has motivated Zhang (1994, 1995) to develop a perturbation
approach for analytical description of the equatorially attached
convection.  Recently this approach has been extended and simplified
by Busse and Simitev (2004). Here we shall just present a short
introduction to the subject. 
\begin{figure}[t]
\begin{center}
\hspace*{-1cm}\epsfig{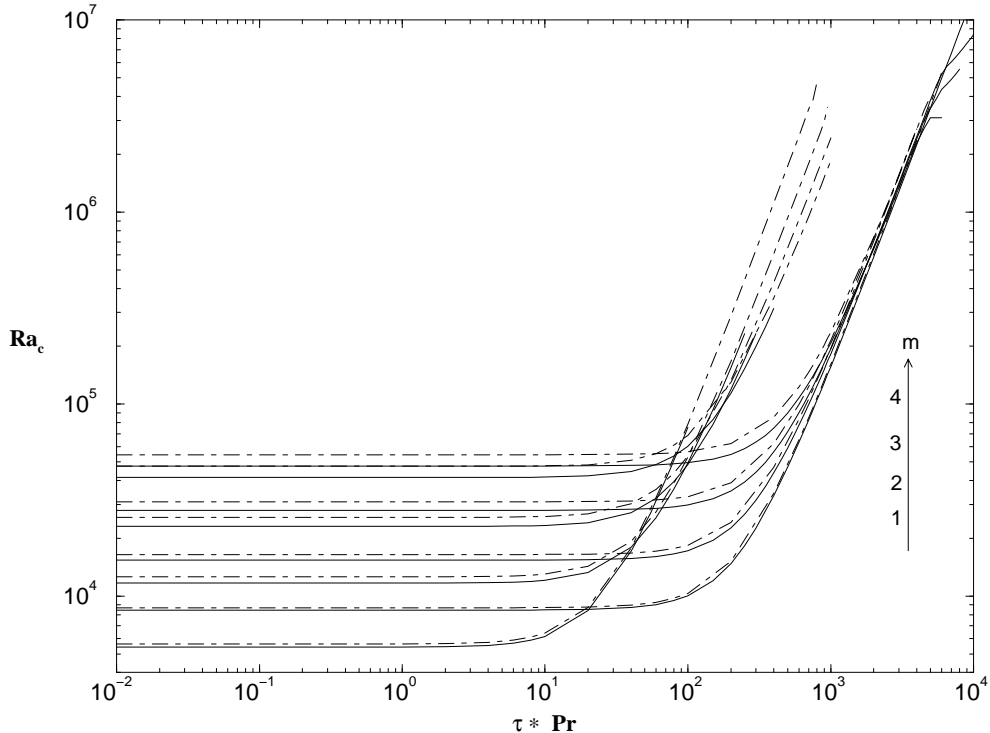}
\end{center}
\caption{The critical Rayleigh number $Ra_c$ (solid lines) for the
  onset of the inertial  convection as a function of  $\tau Pr$ for
  various wavenumbers $m$. The dash-dotted lines correspond to the
  analytical expressions derived in Busse and Simitev (2004). For each
  $m$ there are two modes corresponding to the two signs of $\omega$
  in expression (17b). The retrograde mode corresponding  to the
  positive sign is preferred at lower values of $\tau Pr$. But it
  turns up before the prograde mode which thus is preferred for larger
  values of $\tau Pr$.}
\label{nf8}
\end{figure}

It is obvious from equations (11) even in their linearized versions
that solutions for which the $r-$ and $\theta-$dependences separate
are not admissible in general. Nevertheless for some parameter regimes
simple, physically realistic solutions of the linearized version of
Eqs. (11) together with conditions (13) can be obtained through the
separation ansatz 
\begin{gather}
v = P_m^m( \cos \theta ) \exp \{ im \varphi + i \omega t \} f (r) ,
\nonumber \\ 
w = P_{m+1}^m ( \cos \theta ) \exp \{ i m \varphi + i \omega t \} g
(r) , \nonumber \\
\Theta = P_m^m ( \cos \theta ) \exp \{ im\varphi + i \omega t \} h (r).  
\end{gather}
Solutions of this form satisfy Eqs. (11) after the right hand sides have 
been dropped when the term proportional to $P_{m+2}^m$ in the
expression for $Qw$ can be neglected. This term vanishes exactly in
the limit of small $Pr$ and high $\tau$ when the convection assumes the
form of an inertial wave with the property 
\begin{subequations}
\begin{align}
&f ( r) = \left(\frac{r}{r_o}\right)^m - \left(\frac{r}{r_o}\right)^{m+2}
, \nonumber \\
&g (r) =  \frac{2im(m+2)(r/r_0)^{m+1}}{(2m+1)(\omega (m^2+3m+2)-m)r_o} , \\
&\omega = \frac{\tau }{m+2} ( 1 \pm (1+m(m+2)(2m+3)^{-1})^{\frac{1}{2}} ).
\end{align}
\end{subequations}

Since this solution does not satisfy all the boundary conditions (13),
weak Ekman layers must be added and finite critical Rayleigh numbers for
onset  have thus been obtained (Zhang, 1994). Results for the Rayleigh
number for different values of the azimuthal wavenumber $m$ are shown
in figure \ref{nf8}. According to these results the mode with $m=1$ is
 always preferred in the case $\eta=0$ if $Pr tau$ is sufficiently low
 (Busse and Simitev, 2004). As $Pr tau$ increases transition occurs to
 the mode propagating in the prograde direction.
 Besides the equatorially wall attached mode,
convection can be described approximately  in the form (16) if $\tau$
is less than $10^3$ (see Ardes et al., 1997) or if the thin shell
limit $\eta \rightarrow 1$ is  approached (Busse, 1970b,1973).
\begin{figure}[t]
\begin{center}
\hspace*{0cm}\epsfig{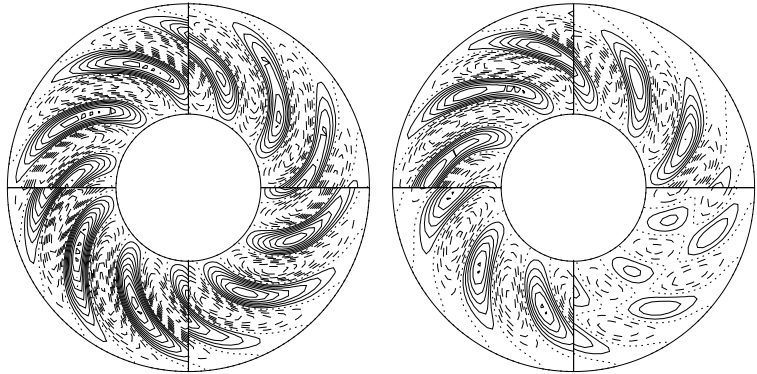}
\end{center}
\caption{Time periodic vacillations of convection at $Ra= 2.8 \cdot
  10^5$ (left side) and $Ra=3 \cdot 10^5$ (right side) for $\tau =
  10^4$, $Pr=1$ The streamlines, $r \partial v / \partial \varphi =$
  const. are shown in one quarter of the equatorial plane. The four
  quarters are equidistant in time (with $\Delta t =0.015$ $(\Delta t
  =  0.024)$ in the left (right) case in the clockwise sense such
  that  approximately a full period is covered by the circles.}
\label{f8}
\end{figure}


\section{ Evolution of Convection Columns at Moderate Prandtl Numbers}

\begin{figure}[t]
\begin{center}
\hspace*{0cm}\epsfig{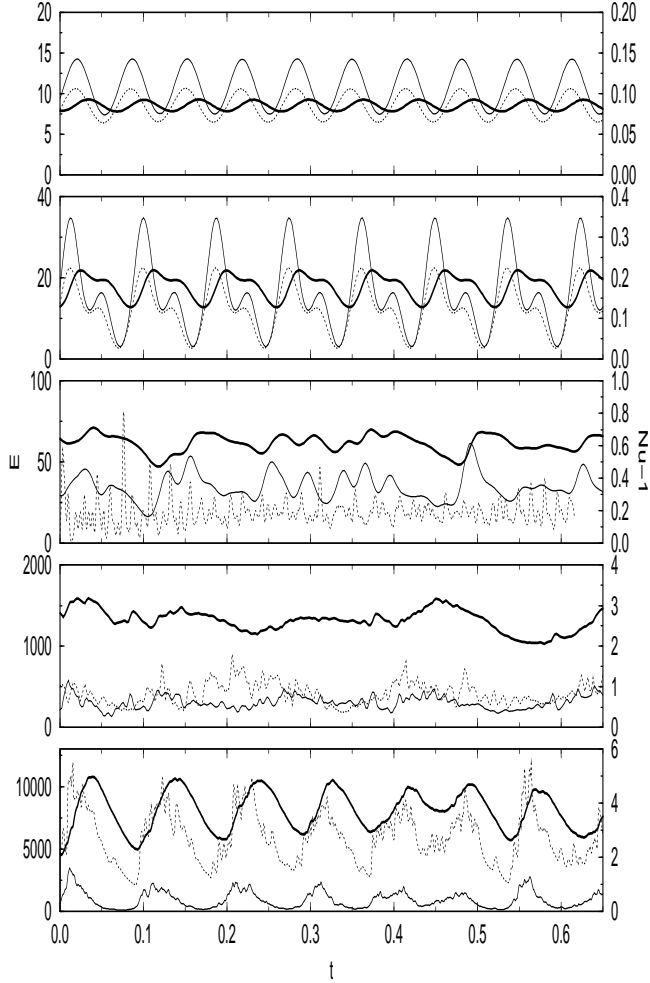}
\end{center}
\caption{Time series of energy densities $E^m_t$ (thick solid
  lines), $ E^f_t$ (thin solid lines) and Nusselt number (dotted
  lines, right ordinate) are  plotted for $Pr=1$, $\tau=10^4$, and
  $Ra=2.8 \cdot 10^5$, $3 \cdot 10^5$, $3.5 \cdot 10^5$, $7 \cdot
  10^5$  and $12 \cdot 10^5$ (from top to bottom). $E^m_p$ and
  $E^f_p$ have not been plotted. $E^m_p$ is several orders of
  magnitude smaller than the other energies and $ E^f_p$ always
  approaches closely $0.4 \cdot E^f_t$.} 
\label{f9}
\end{figure}
In general the onset of convection in rotating fluid spheres occurs
supercritically. As long as the convection assumes the form of shape
preserving travelling thermal Rossby waves as described by linear theory,
its azimuthally averaged properties are time independent. In fact, as seen
from a frame of reference drifting together with the convection columns
the entire pattern is steady. A differential rotation is generated
through the action of the Reynolds stress as explained in section
2. The latter is caused by the spiralling 
cross section of the columns which persists as a dominant feature at
moderate Prandtl numbers far into the turbulent regime. The plots of the
streamlines $r \frac{\partial v}{\partial \phi}= \mbox{const.}$ in the
equatorial plane shown in any of the quarter circles of figure
\ref{f8} give  a good impression of the spiralling nature of the columns. 
\begin{figure}[t]
\begin{center}
\hspace*{0cm}\epsfig{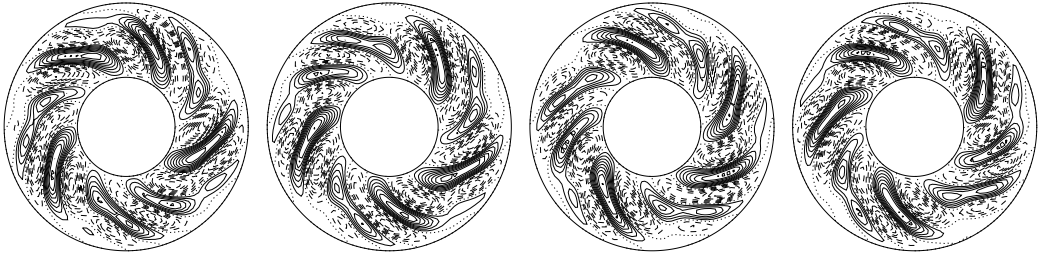}
\end{center}
\caption{Modulated shape vacillations of convection for $Ra=2.9 \cdot
  10^5$, $\tau=10^4$, $Pr=1$. The plots show streamlines,
  $r \frac{\partial v}{\partial \phi}=const.$, in the equatorial
  plane and are equidistant in time with $\Delta t =0.04$ so that
  approximately a full period is covered.} 
\label{f10}
\end{figure}

A true time dependence of convection develops in the form of vacillations
after a subsequent bifurcation. First the transition to amplitude
vacillations occurs in which case just the amplitude of convection varies
periodically in time as exhibited in the left plot of figure \ref{f8}. At a
somewhat higher Rayleigh number shape vacillations become noticeable
which are characterized by periodic changes in the structure of the
columns as shown in the right plot of figure \ref{f8}. The outer part of the
columns is stretched out, breaks off and decays. The tendency
towards breakup is caused by the fact that the local frequency of
propagation varies with distance from the axis according to expression
(15c) after $\theta_m$ has been replaced by the local colatitude $\theta$.

The two types of vacillations also differ significantly in their
frequencies of oscillation. This is evident from the time records of the
energy densities of convection which have been plotted in figure
\ref{f9}. This figure gives an overview of the evolution of time
dependence in the interval $2.8 \cdot 10^5 \leq Ra \leq 10^6$.
The various components of the energy densities are defined by
\begin{subequations}
\begin{align}
\label{7}
E_p^m = \frac{1}{2} \langle \mid \nabla \times ( \nabla \bar v \times \vec r 
) \mid^2 \rangle , \quad & E_t^m = \frac{1}{2} \langle \mid \nabla \bar w \times
\vec r \mid^2 \rangle \\
\label{8}
E_p^f = \frac{1}{2} \langle \mid \nabla \times ( \nabla \check v \times \vec r) 
\mid^2 \rangle , \quad&  E_t^f = \frac{1}{2} \langle \mid \nabla \check w \times
\vec r \mid^2 \rangle
\end{align}
\end{subequations}
where $\bar v$ refers to the azimuthally averaged
component of $v$ and $\check v$ is given by $\check v = v - \bar v $. 
\begin{figure}[t]
\begin{center}
\hspace*{0cm}\epsfig{file=FINAL.ps,height=9cm,bbllx=96,bblly=76,bburx=567,bbury=744,angle=-90,clip=}
\end{center}
\caption{Energy densities $E^m_t$ (solid line), $E^m_p$ (thin
  dashed line,  multiplied by the factor 100), $E^f_t$  (short
  dash - long dash line), $E^f_p$ (thick dashed line) and the
  Nusselt number $Nu$ (dotted line, right ordinate) are plotted as
  function of $Ra-Ra_c$ in the case $\tau = 10^ 4$, $Pr =1$.  $Ra_c =
  1.9 \cdot 10^5$ has been used corresponding to $m=10$} 
\label{f12}
\end{figure}

With a further increase of the Rayleigh number spatial modulations of
the shape vacillations occur as shown in figure \ref{f10}. These
modulations often correspond to a doubling of the azimuthal period but
soon contributions 
with the azimuthal wavenumber $m = 1$ arise as shown in figure
\ref{f10}.  The pattern in this particular case is still periodic if
appropriately sifted in azimuth. But as further modulations enter
convection becomes quasiperiodic and 
with increasing $Ra$ a chaotic state is reached. We refer to the
recent paper of Simitev and Busse (2003) which includes some movies to
demonstrate the time dependence of convection. Figure \ref{f9} also
demonstrates the diminishing fraction of the total kinetic energy that
is associated with the poloidal component of motion which carries the
convective heat transport. The differential rotation 
in particular increases much faster with $Ra$ than the amplitude of convection
since the Reynolds  stress is proportional to the square of the
latter. While this is already evident from the sequence of    plots in
figure \ref{f9} it is even more obvious from figure
\ref{f12}. Here  it can also be seen that the onset of vacillations
and aperiodic time dependence tends to increase the heat transport as
indicated  by the Nusselt number in contrast to the situation in a planar
convection layer with the same    Prandtl number (Clever and Busse,
1987). In the latter case the mismatch is absent between the structure
of the convection flow and the configuration of the boundary which
inhibits the heat transport in rotating spherical fluid shells. The
time varying shift in the radial position of the convection columns
thus promotes the heat transport.  
\begin{figure}[t]
\begin{center}
\hspace*{0cm}\epsfig{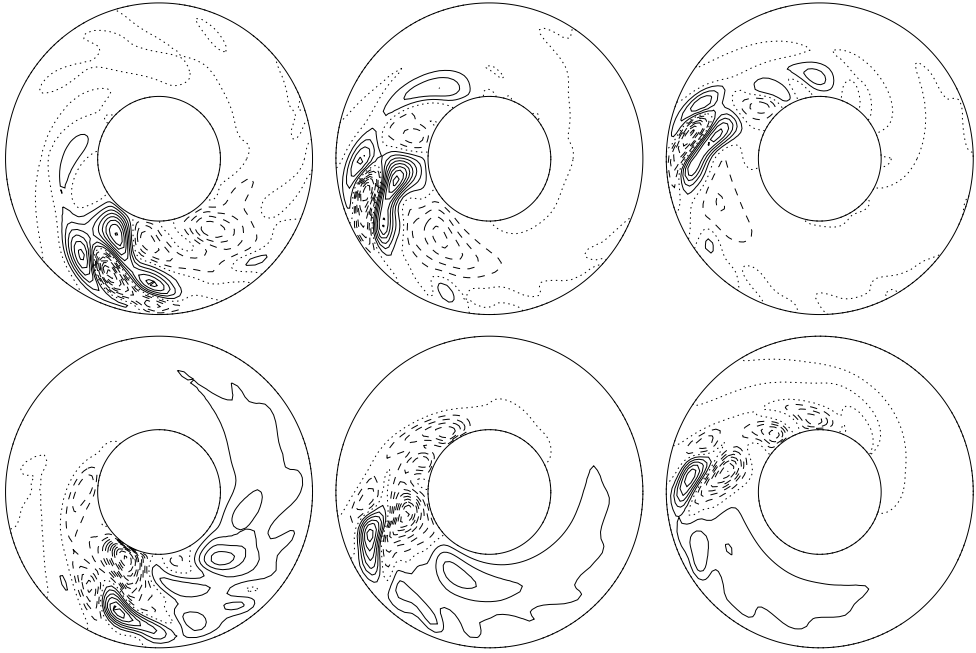}
\end{center}
\caption{Localized convection for $Ra=7 \cdot 10^5, \tau =1.5 \cdot 10^4, Pr=0.5$
The streamlines, $r \partial v / \partial \varphi =$ const. (first row) and the
isotherms, $\Theta =$ const. (second row), are shown in the equatorial plane for
equidistant times (from left to right) with $\Delta t = 0.03$.}
\label{f13}
\end{figure}
 
\begin{figure}[t]
\begin{center}
\hspace*{0cm}\epsfig{file=fig14.epsi,width=6cm,height=10cm,angle=-90}
\end{center}
\caption{Relaxation oscillation of turbulent convection in the case
  $Ra= 10^6$, $\tau = 1.5 \cdot 10^4$, $Pr=0.5$. Energy densities $\bar E_t$
  (solid line), $\check E_t$ (dotted line),  $\check E_p$ (short
  dashed line) and the Nusselt number (long dashed line, right
  ordinate) are shown as functions of time $t$. }
\label{f14}
\vspace{0.5cm}
\hspace*{0cm}\epsfig{file=fig15.epsi,width=13.8cm,angle=0}
\caption{Sequence of plots starting at $t=0.12015$ and equidistant in
  time $(\Delta t=0.016$) for 
  the same case as in Fig.~\ref{f14}.   Lines of constant
  $\bar{u_{\varphi}}$ and mean temperature perturbation,
  $\bar{\Theta}=$ const. in the meridional plane, are shown in  the left 
  and right halves, respectively, of the first row. The second 
  row show streamlines, $r \partial v / \partial \varphi =$
  const., in the equatorial plane.}
\label{f15}
\end{figure}
Surprisingly the spatio-temporal randomness of convection columns does not
just increase at larger values of $Ra$, but instead new coherent structures
evolve (Grote and Busse, 2001).
First there is localized convection as shown in Fig.~\ref{f13}.
The differential rotation has become so strong that its shearing action 
inhibits convection in most parts of the spherical fluid shell as is evident 
from the figure.
Only in a certain region of longitude is convection strong enough to overcome
the shearing action of differential rotation.
In the ``quiet''zone the basic temperature profile recovers towards the purely
conducting state and thus provides the buoyancy in the interior of the
shell which sustains the 
localized convection as  it is  recirculated into the ``active''
zone by the differential rotation.

After a further amplification of the differential rotation with increasing
$Ra$ the local intensification of convection no longer suffices to 
overcome the shearing action of the zonal flow.
Instead of a spatial separation between ``active'' and ``quiet'' zones
the system chooses a separation in time
which manifests itself in the relaxation oscillations seen in the
lowermost plot of  Fig.~\ref{f9}.
The fluctuating component of motion is still rather turbulent in the case
of the relaxation oscillation as demonstrated in Figs.~\ref{f14} and \ref{f15}.
When the differential rotation has decayed sufficiently in the near absence
of Reynolds stresses generated by convection, a sudden burst of convection
activity occurs leading to a sharp peak in the heat transport. But since the
Reynolds stress grows just as suddenly as the kinetic energy of convection,
the growth of the differential rotation occurs with only a slight
delay. The shearing off of the  convection columns then leads to their
decay  almost as quickly as they had  set in.
The relaxation oscillations occur over a wide region in the parameter space
for high Rayleigh and Coriolis numbers and for Prandtl numbers of the order
unity or less. Since it is mainly determined by the viscous decay of
the differential  rotation, the period of the relaxation oscillation does not vary much with these
parameters. For the case of $\eta=0.4$ which has been used for almost
all numerical simulations a period of about 0.1 is usually found.
\begin{figure}[t]
\begin{center}
\hspace*{0cm}\epsfig{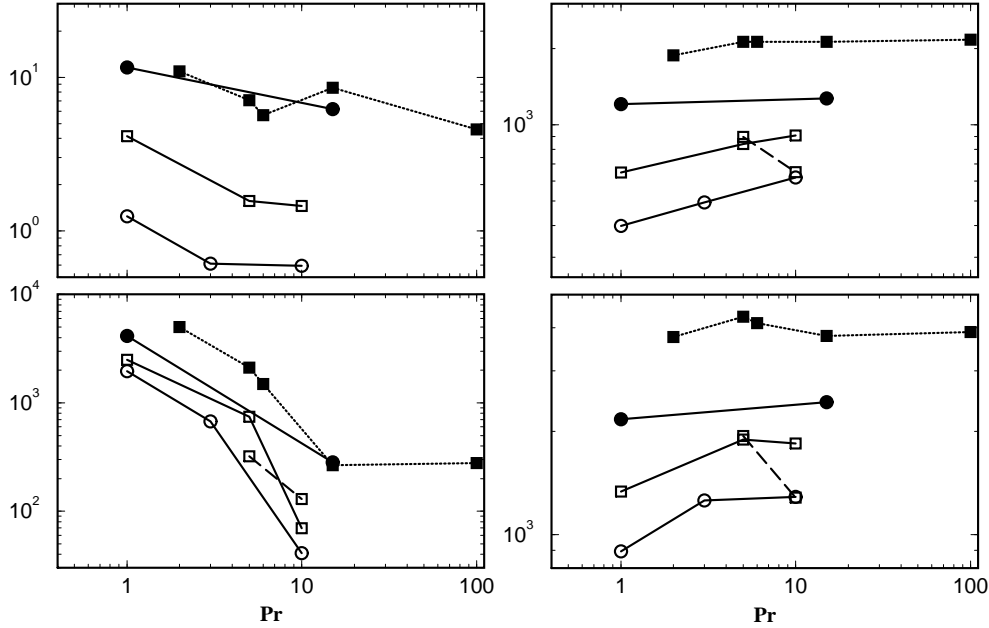}
\end{center}
\caption{Kinetic energy densities $E^m_p$ (upper left panel), $E^f_p$
  (upper right panel), $E^m_t$ (lower left panel) and $E^f_t$ (lower
  right panel) all multiplied by $Pr^2$ as a function of the Prandtl
  number $Pr$ in the case $\tau=5 \cdot 10^3$. 
  The values of the Rayleigh number $R = 5\cdot10^5$, $6\cdot10^5$,
  $8\cdot10^5$, $10^6$ are denoted by empty circles and squares and
  full circles and squares, respectively.}
\label{f16}
\end{figure}


\section{Finite Amplitude Convection at Higher \\
Prandtl Numbers}

\begin{figure}[t]
\begin{center}
\hspace*{0cm}\epsfig{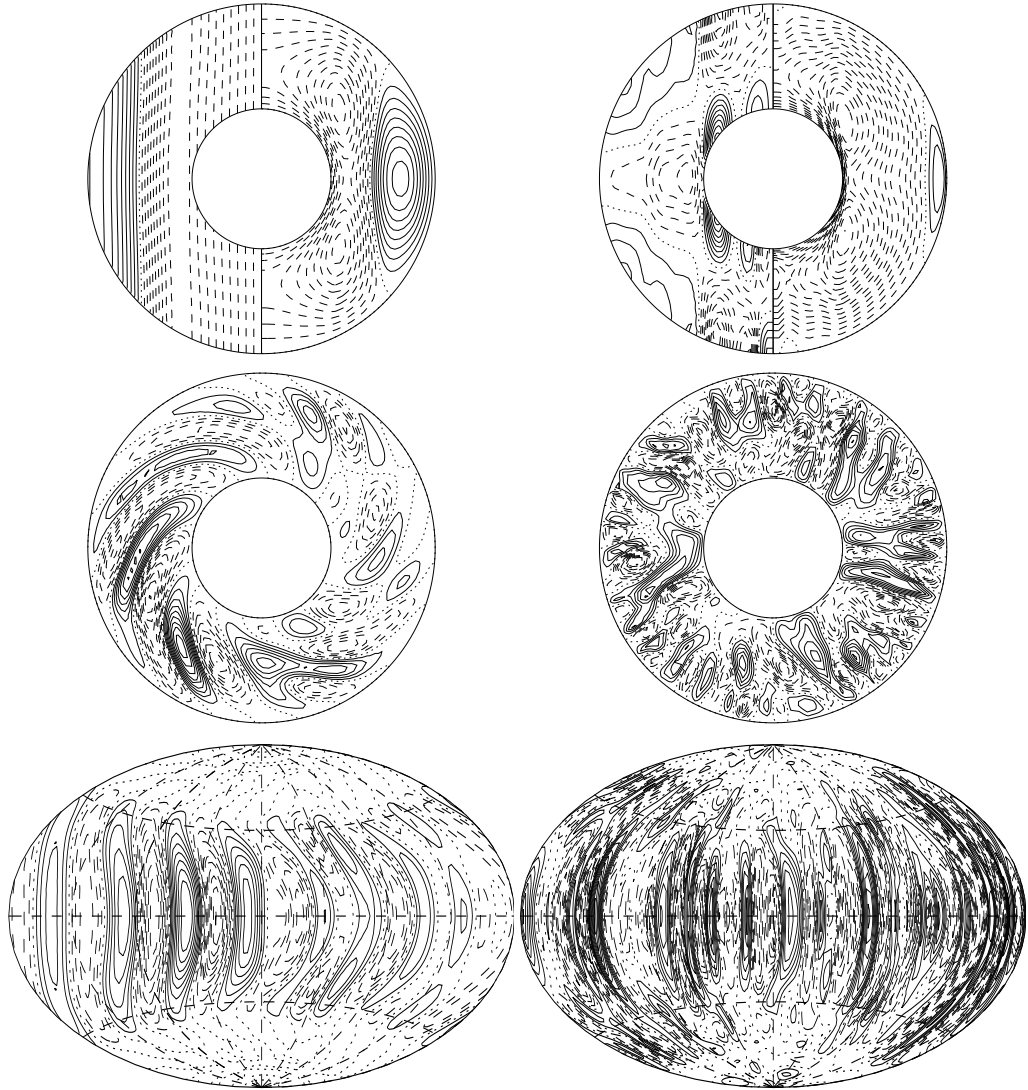}
\end{center}
\caption{Convection in rotating spherical fluid shells in the cases
  $\tau = 10^4$, $Ra=4 \cdot 10^5$, $Pr=1$ (left column) and $\tau = 5
  \cdot 10^3$, $Ra= 8 \cdot 10^5$, $Pr=20$  (right column). Lines of
  constant mean azimuthal velocity $\bar u_{\varphi}$ are shown in the
  left halves of the upper circles and isotherms of $\bar{\Theta}$
  are shown in the right halves. The plots of the middle row show
  streamlines, $r \partial v / \partial \varphi=$ const., in the
  equatorial plane. The lowermost plots indicate lines of constant
  $u_r$ in the middle spherical surface, $r=r_i +0.5$.}
\label{f17}
\end{figure}

The transitions from drifting convection columns to vacillating
convection and modulated vacillating convection do not change much as
as the Prandtl number tends to high values (Zhang, 1992b). But as $Pr$
increases the influence of the differential
rotation which dominates the evolution of the convection columns for
Prandtl numbers  of the order unity and below diminishes rapidly. The
feedback process exhibited most clearly by the mean flow instability
discussed in section 2 ceases to operate at Prandtl numbers of the
order 10. Above this value of $Pr$ the properties of
convection become nearly independent of $Pr$ when the thermal time scale
instead of the viscous one is used. This property which is familiar
from Rayleigh-B\`enard  convection in plane layers heated from below
holds rather generally in convection systems. For this reason we have
plotted in figure \ref{f16} energy densities as defined by expressions
(18), but  multiplied by the factor $Pr^2$ in order to demonstrate the
tendency towards independence of $Pr$.  It should be noted that most
of values used in this figure have been obtained from dynamo
computations. But the action of the Lorentz force is rather weak and
hardly affects the Prandtl number dependence  of convection. Only the
lower left plot for the energy densities of the differential rotation
shows the expected strong decay with increasing $Pr$. These energy
densities  do not decay to zero for $Pr \rightarrow 0$, however. The
thermal wind relationship obtained from the azimuthal average of the
$\phi$-component of the curl of equation (10a)  
\begin{equation}
\tau {\bf k}\cdot \nabla \bar u_\phi = \partial _\theta \bar \Theta
\end{equation}
continues to require a finite field $\bar u_\phi$ in the limit of high
$Pr$. The right hand side is finite because the azimuthally  averaged
temperature field $\bar \Theta$ deviates strongly from spherical
symmetry as long as the influence of rotation is significant.

The typical differences between convection at Prandtl numbers of the
order unity and higher values are exhibited in figure \ref{f17} where
typical results obtained for $Pr = 1$ and $Pr = 20$ are compared at
similar values of $Ra$ and $\tau$. The approximate validity of the
thermal wind 
relationship (19) can be noticed from a comparison of the plots of
$\bar u_\phi$ and of  $\bar \Theta$ in the case $Pr = 20$. The
convection columns retain their alignment with the axis of rotation
with increasing $Pr$, but the spiralling nature of their radial
orientation disappears. 

\section{ Finite Amplitude Inertial Convection}

\begin{figure}[t]
\begin{center}
\hspace*{0cm}\epsfig{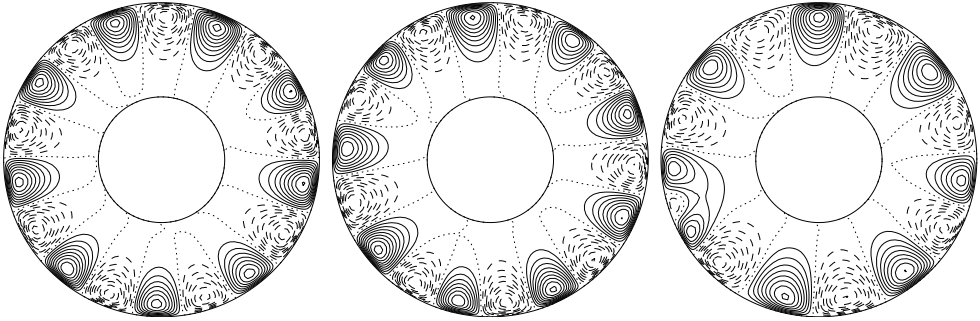}
\end{center}
\caption{Streamlines $r \frac{\partial v}{\partial \phi}=const.$ in
the equatorial plane for the case $Pr=0.025$, $\tau=10^5$ with $Ra=3.2
\cdot 10^5$, $3.4 \cdot 10^5$, $4 \cdot 10^5$ (from left to right).}
\label{f18}
\end{figure}

\begin{figure}[t]
\begin{center}
\hspace*{-0.4cm}\epsfig{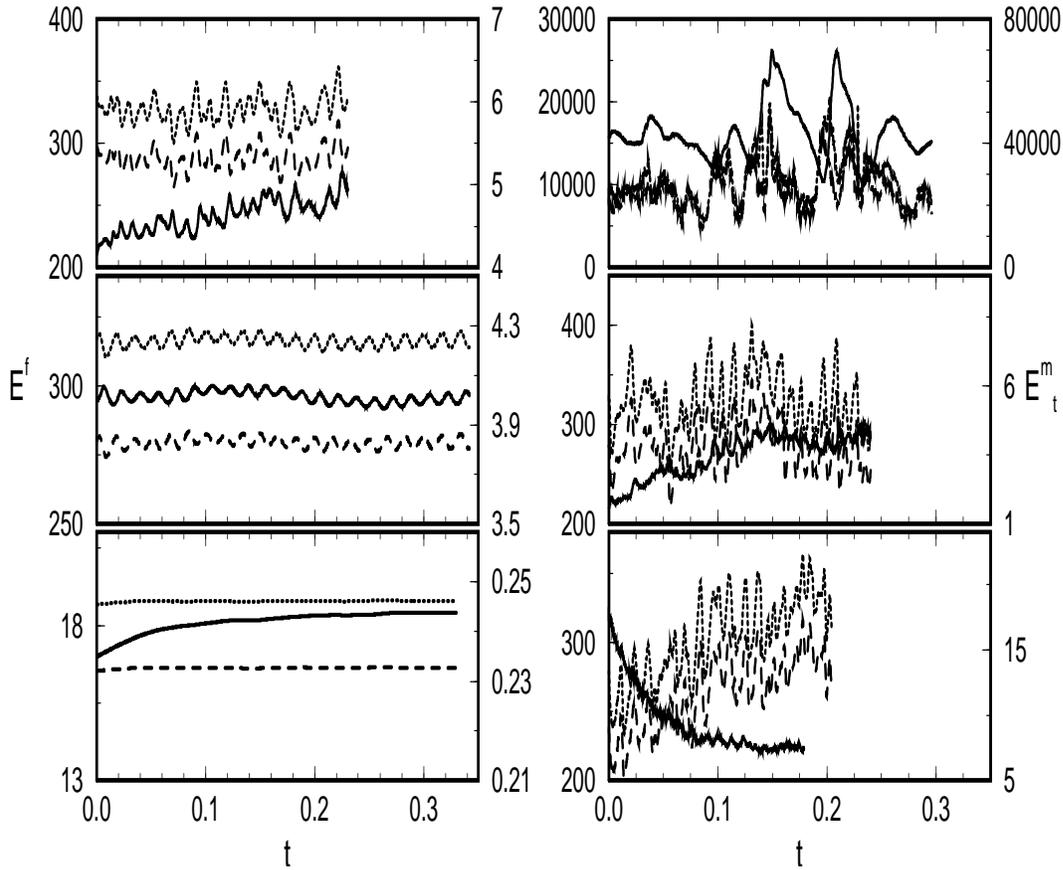}
\end{center}
\caption{Time series of energy densities of convection in the case
  $Pr=0.025$, $\tau=10^5$, for $Ra=3.1 \cdot 10^5$, $3.2 \cdot 10^5$,
  $3.4 \cdot 10^5$, (from bottom to top, left side) and $3.8 \cdot
  10^5$, $4 \cdot 10^5$, $8 \cdot  10^5$  (from bottom to top, right
  side).  The solid, dotted and dashed lines correspond to $E^m_t$,
  $E^f_t$, $E^f_p$ , respectively. $E^m_t$ is measured on the right
  ordinate.}   
\label{f19}
\end{figure}
For an analysis of nonlinear properties of equatorially attached
convection we focus on the case $Pr=0.025$ with $\tau=10^5$. The
critical Rayleigh number for this case is $Ra_c=28300$ corresponding to
$m=10$. As $Ra$ is increased beyond the critical value other values of
$m$ from 7 to 12 can be realized, but $m=10$ and lower values are
usually preferred. An asymptotic perfectly periodic solution with
$m=10$ or $m=9$ can be found only for Rayleigh numbers close to the
critical value when computations are started from arbitrary initial
conditions. On the other hand, perfect periodic patterns appear to be
stable with respect to small disturbances over a more extended regime
of supercritical Rayleigh numbers. Distinct transitions like the
transition to amplitude vacillations and to structure vacillations do
not seem to exist for equatorially attached convection. Instead
modulated patterns are typically already observed when $Ra$ exceeds the
critical value by 10\% as can be seen in the plots of figure
\ref{f18}. These modulations are basically caused by the superposition
of several modes with neighboring values of the azimuthal wavenumber $m$
which appear to propagate nearly independently. For example, the
period of $1.68 \cdot 10^{-2}$ visible in the energy densities shown
in figure \ref{f19} in the case $Ra=3.2 \cdot10^5$ is just about half the
difference of $3.34 \cdot 10^{-2}$  between the periods of the modes
$m=8$ and $m=9$ according 
\begin{figure}[t]
\begin{center}
\hspace*{-0.4cm}\epsfig{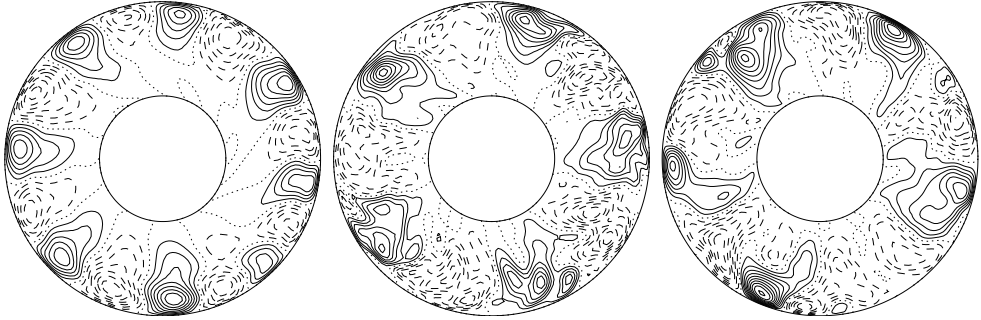}
\end{center}
\caption{Streamlines $r \frac{\partial v}{\partial \phi}=const.$ in
the equatorial plane for the case $Pr=0.025$, $\tau=10^5$ with $Ra=6
\cdot 10^5$, $8 \cdot 10^5$, $10^6$ (from left to right).}
\label{f20}
\end{figure}
the inertial wave dispersion relation (17b) for prograde modes. This
property must be expected if a small component with $m=8$ participates
in the pattern shown in figure \ref{f18} which is dominated by the
($m=9$)-component.  The time series of energy densities shown
in figure \ref{f19} indicate that usually more   than two modes 
contribute to the dynamics of the pattern with the exception of the
case just discussed since the time dependence is
not periodic as in the case when only two modes interact. The
computations of the time series require a high spatial resolution
together with a small time step. The time spans indicated in figure
\ref{f19} are sufficient for reaching a statistically steady state of the
fluctuating components of motion since these equilibrate on the fast
thermal time scale of the order $Pr^{-1}$. Only close to $Ra_c$ the
adjustment process takes longer as can be seen in the case $Ra=3.1
\cdot 10^5$ where a ($m=10$)-pattern approaches its equilibrium
state. The pattern corresponding to the other cases of figure
\ref{f19} are shown in figures \ref{f18}, and
\ref{f20}. The  differential rotation represented by $E_t^m$ relaxes
on the viscous time scale and therefore takes a long time to reach its
asymptotic regime in the examples shown in figure \ref{f19}. But the
differential rotation is quite weak such that it has a negligible
effect on the other components of  motion except in the case of the
highest Rayleigh number of figure \ref{f19}. Even smaller is the
axisymmetric part of the poloidal component of  motion which is not
shown in the plots of figure \ref{f19}.  At higher Rayleigh numbers
the convection eddies spread  farther into the interior and in
some cases become detached from the equator as can be seen in the
plots of figure \ref{f20}. In this way the convection eddies contribute to
the heat transport from the inner boundary. But at the same time they
acquire the properties of the convection columns which are
characteristic for convection at higher Prandtl numbers. Accordingly
the differential rotation is steeply increased at $Ra=10^6$ and a
tendency towards relaxation oscillation can be noticed in the upper
right time series of figure \ref{f19}. For additional details on low
Prandtl number spherical convection we refer to Simitev and Busse (2003).

\end{document}